\documentclass[hidelinks]{article}

\usepackage{cite}
\usepackage{amsmath,amssymb,amsfonts}
\usepackage{algorithmic}
\usepackage{graphicx}
\usepackage{textcomp}
\usepackage{xcolor}
\usepackage{bm}
\usepackage{orcidlink}
\usepackage[misc]{ifsym}
\usepackage{makecell}
\usepackage{listings}
\usepackage{stfloats}
\usepackage{rotating}
\usepackage{hyperref}
\usepackage{authblk}

\newcommand{\comp}[3]{#1\, \Rrightarrow\, #2,\, #3}

\newcommand{\keywords}[1]{%
    \vspace{2ex}%
    \noindent\textbf{Keywords: }%
    \hangindent=2em\hangafter=1 #1%
    \par\vspace{2ex}%
}

\lstdefinelanguage{mimosa}
{
    morekeywords={step,node,channel,implements,every,ms,pre,fby,if,then,else,Some,None},
    sensitive=false,
    morecomment=[s]{(*}{*)},
}

\date{}

\begin{document}

\author[1]{Nikolaus Huber\,\orcidlink{0000-0002-1616-0602}\,}
\author[1,2]{Susanne Graf\,\orcidlink{0000-0003-4354-6807}\,}
\author[1,3]{Philipp Rümmer\,\orcidlink{0000-0002-2733-7098}\,}
\author[1]{Wang Yi\,\orcidlink{0000-0002-2994-6110}\,} 
\affil[1]{Uppsala University, Uppsala, Sweden}
\affil[2]{Univ. Grenoble Alpes, CNRS, Grenoble INP, VERIMAG, France}
\affil[3]{University of Regensburg, Regensburg, Germany}

\title{Compiling the Mimosa programming language\\ to RTOS tasks}

\maketitle

\begin{abstract}
  \noindent This paper introduces a compilation scheme for programs written in the Mimosa programming language, which builds upon the MIMOS model of computation. Mimosa describes embedded systems software as a collection of time-triggered processes which communicate through FIFO queues. We formally describe an adaptation of the Lustre compilation scheme to the semantics of Mimosa and show how the coordination layer can be mapped to real-time operating system primitives.
\end{abstract}

\section{Introduction}
\label{sec:introduction}

MIMOS~\cite{wang-mimos} is a new model of computation for asynchronous implementation of embedded systems software. It builds upon the well-known concept of Kahn process networks~\cite{kahn-networks} (KPNs), where software is described through nodes which communicate exclusively via (conceptually unbounded) FIFO queues. MIMOS extends this model by assigning a periodic release pattern to each node and defining reading and writing strategies for each port. At the beginning of a period, a node checks if sufficient data is available according to its reading strategy, then performs its computation and writes the results at the end of the period. If there is not enough input available, the node stays idle until its next release time.

\begin{figure}[ht]
    \centering
    \includegraphics[width=0.6\linewidth]{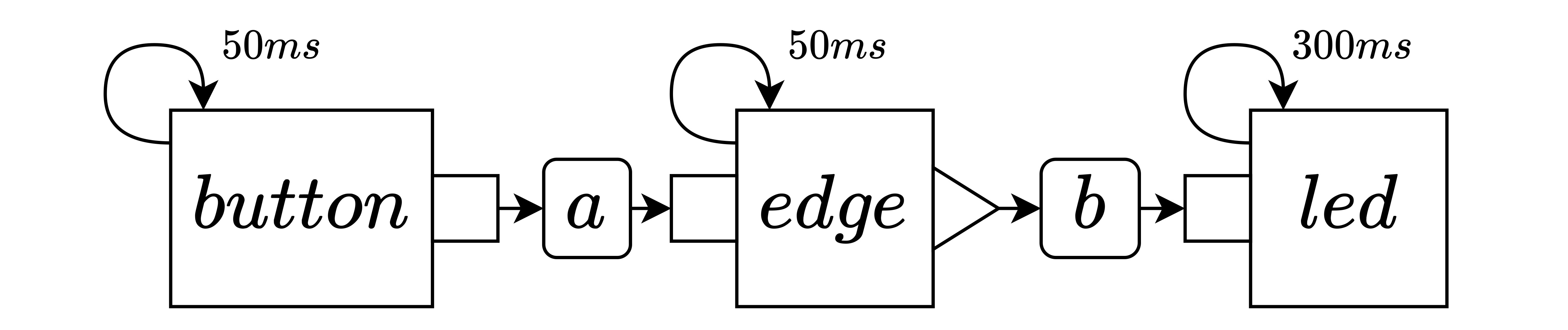}
    \caption{Edge detector example}
    \label{fig:example-edge-detect-graphical}
\end{figure}

Figure~\ref{fig:example-edge-detect-graphical} illustrates an example of a MIMOS network. There are three nodes, \texttt{button}, \texttt{edge}, and \texttt{led}. The overall aim is to toggle an LED whenever the button is pressed. Node \texttt{button} polls the current pin level, \texttt{edge} detects rising and falling edges in the signal, and \texttt{led} performs the toggle on each received rising edge. The first two nodes run with the same period of $50\,ms$, while the LED handler runs with a lower frequency. This would inevitably fill up the buffer $b$. Therefore, \texttt{edge} only outputs a value whenever an actual edge is detected. If we know that edges only appear infrequently, the last node can be run with a higher period. This \emph{optional output} behaviour is indicated through a triangle output port on the middle node.

In our previous work~\cite{huber-mimosa}, we have introduced a prototype programming language on top of MIMOS, called Mimosa. It provides the primitives to implement all components of a MIMOS network, including \emph{steps} to describe computation, \emph{channels} for communication, and \emph{nodes} for describing periodically triggered processes. In~\cite{huber-mimosa}, only a simulator for the semantics of Mimosa has been presented. Our main contribution in this paper is the description of a compilation scheme for Mimosa programs to C code, which executes on top of a real-time operating system (RTOS). This represents a significant step forward in running Mimosa on actual embedded hardware.

The compilation of steps to sequential code presented in this paper is an adaptation of the compilation scheme for the Lustre language proposed in~\cite{biernacki-clock-directed-compilation}. Compared to Lustre, Mimosa employs demand-driven evaluation for certain sub-expressions, which enables the handling of side-effectful computations more conveniently without relying on multiple clocks. For example, in the \texttt{led} node in Figure~\ref{fig:example-edge-detect-graphical}, the LED shall only be toggled on rising edges, which a simple conditional expression can express in Mimosa (see Section~\ref{sec:mimosa-lang}). Our adaptation of the Lustre compilation scheme, therefore, keeps the internal block structure of specific operations and removes multi-clock support.

This paper is organised in the following way: Section~\ref{sec:mimosa-lang} illustrates actual Mimosa code using the example from Figure~\ref{fig:example-edge-detect-graphical}, and it also explains the different language elements further. Section~\ref{sec:compilation} describes the compilation routine, which is divided into two steps: compiling the sequential code and translating the nodes and channels into RTOS primitives. We give an overview of related work in Section~\ref{sec:related-work} and outline future work in Section~\ref{sec:conclusion}.
\section{A Mimosa example}
\label{sec:mimosa-lang}

Syntactically, Mimosa builds upon the Lustre~\cite{caspi-lustre} data-flow language, from which it retains the equation-based definition of computation. Other than Lustre, expression evaluation is not assumed to be side-effect-free, which forces a different evaluation strategy. For example, while the conditional expression (i.e. \textbf{if/then/else}) is defined as a point-wise combination of the (infinite) \emph{flows} of values generated from the two branch sub-expressions in Lustre, they are only selectively evaluated in Mimosa.

\begin{figure}[ht]
\begin{lstlisting}[
    language=mimosa, 
    basicstyle=\footnotesize, 
    numbers=left,
    numbersep=5pt,
    numberstyle=\tiny
]
  step poll () --> (_ : bool)
  step toggle_led () --> ()

  step edge (in : bool) --> (out : bool?)
  {
    pre_in = in -> pre in;
    out = if !pre_in && in then Some true
          else if pre_in && !in then Some false
          else None;
  }

  step toggle (in : bool) --> ()
  {
    _ = if in then toggle_led () else ();
  }

  channel a : bool
  channel b : bool

  node button implements poll () --> (a) every 50ms
  node edge implements edge (a) --> (b?) every 50ms
  node led implements toggle (b) --> () every 300ms
\end{lstlisting}
\caption{Edge detector in Mimosa}
\label{fig:edge-detector-mimosa}
\end{figure}

Figure~\ref{fig:edge-detector-mimosa} shows the textual Mimosa program for the example in Figure~\ref{fig:example-edge-detect-graphical}, which consists of a set of top-level definitions including \emph{steps}, \emph{channels}, and \emph{nodes}. A step (Mimosa's equivalent to a function definition) consists of a \emph{step signature} defining a \emph{name} and \emph{input} and \emph{output patterns} (i.e. possibly nested lists of name bindings), optionally followed by a list of \emph{sequence equations}, each defined by a \emph{pattern} and an \emph{expression}. Expressions are formed inductively from sub-expressions, their (abstract) syntax is further described in Section~\ref{sec:compilation}. Step signatures without a body are called \emph{prototypes}, their implementation must be provided externally.

In the example, four steps are defined, two of which are prototypes. The step \texttt{edge} detects rising and falling edges in the Boolean input signal. Its output is defined as an \emph{optional Boolean} (indicated by the type annotation \texttt{bool?}), where \textbf{Some} \texttt{true} indicates a rising edge, \textbf{Some} \texttt{false} a falling edge, and \textbf{None} a stagnant signal. The \texttt{pre\textunderscore in} sequence is defined through the \emph{memory operators} \texttt{->} and \textbf{pre}, which are further described in Section~\ref{sec:compilation}.

The step \texttt{toggle} calls the step \texttt{toggle\textunderscore led} whenever it receives a rising edge as an input. As mentioned before, only one branch is evaluated in a conditional expression in Mimosa. The same equation in Lustre would have caused \texttt{toggle\textunderscore led} to be called in each step invocation, independently of the value of the \texttt{in} argument.

A \emph{channel} denotes a FIFO buffer holding values of a particular type. Even though they are conceptually unbounded, for efficient implementation, their bounds need to be provided during compilation (see Section~\ref{sec:compile-coordination-layer}).

A node is an implementation of a particular step as a \emph{process}. Each node lists its \emph{name}, which step it implements, which channels it reads data from and writes data to, and its \emph{period}. Each node periodically checks whether there is at least one data item in each of its (non-optional) input channels, in which case it calls its implemented step with those items. Otherwise, it waits until its next release. If a node computes, it writes its outputs at the end of the period.

Input and output channels can be marked as \emph{optional} by appending the $?$ operator to the name. An optional input does not block the computation: if a data item $t$ is available in the channel, it is wrapped inside an optional value \textbf{Some} $t$, otherwise \textbf{None} is used. Similarly, if an output is marked as optional, then an output \textbf{Some} $t$ from the step is unwrapped, and only the value $t$ is written to the respective channel. If the output is \textbf{None}, no value is written. In the example in Figure~\ref{fig:edge-detector-mimosa}, the output of the \texttt{edge} node is declared as optional, so channel \texttt{b} only contains the detected edges.
\section{Compilation}
\label{sec:compilation}

The Mimosa language naturally divides into two layers: the \emph{step layer} (i.e., steps and prototypes) and the \emph{coordination layer} (i.e., channels and nodes). We describe the compilation of each of these layers independently.

\subsection{Compiling steps}

\begin{figure}
\small
\begin{align*}
\begin{array}{r@{\quad}c@{\quad}l@{\quad}l}
e : \text{Expression} & ::= & x & \text{Variable} \\
{} & | & c & \text{Constant} \\
{} & | & e\bm{,}\, \ldots\bm{,}\, e & \text{Tuple} \\
{} & | & \textbf{pre}\, e & \text{Pre} \\
{} & | & e\, \bm{\rightarrow}\, e & \text{Initialised-by} \\
{} & | & e\, \textbf{fby}\, e & \text{Followed-by} \\
{} & | & e\, e & \text{Application} \\
{} & | & \textbf{if}\, e\, \textbf{then}\, e\, \textbf{else}\, e & \text{Conditional} \\
{} & | & \textbf{None} & \text{None} \\
{} & | & \textbf{Some}\, e & \text{Some} \\
{} & | & \textbf{either}\, e\, \textbf{or}\, e & \text{Option match} \\
p : \text{Pattern} & ::= & x & \text{Variable pattern} \\
{} & | & p\bm{,}\, \ldots\bm{,}\, p & \text{Tuple pattern}
\end{array}
\end{align*}
\caption{Abstract syntax of Mimosa expressions}
\label{fig:mimosa-abstract-syntax}
\end{figure}

Each step that is not a prototype consists of a list of equations, which are formed from \emph{patterns} and \emph{expressions}. Their abstract syntax is shown in Figure~\ref{fig:mimosa-abstract-syntax}. Compared to the syntax provided in our previous work~\cite{huber-mimosa}, we have omitted function abstraction as an expression, allowing functions to be implemented only at the top level as steps. They could be added as first-class values by first lambda-lifting~\cite{johnsson-lambda-lifting} them to the global scope. 

Expressions are formed inductively from sub-expressions. \emph{Variables}, \emph{constants}, \emph{tuple construction}, \emph{application}, and \emph{conditional execution} are well known from other programming languages. 

\begin{figure}
\small
\centering
\begin{tabular}{c|llll}
\textbf{Cycle} & 1 & 2 & 3 & $\cdots$ \\ 
\hline
$x$ & $x_1$ & $x_2$ & $x_3$ & $\cdots$ \\
$y$ & $y_1$ & $y_2$ & $y_3$ & $\cdots$ \\
$\textbf{pre}\, x$ & $\bot$ & $x_1$ & $x_2$ & $\cdots$ \\
$x\, \bm{\rightarrow}\, y$ & $x_1$ & $y_2$ & $y_3$ & $\cdots$ \\
$x\, \textbf{fby}\, y$ & $x_1$ & $y_1$ & $y_2$ & $\cdots$
\end{tabular}
\caption{Memory operators}
\label{fig:memory-operators}
\end{figure}

The \emph{memory operators} (\textbf{pre}, $\bm{\rightarrow}$, and \textbf{fby}) are defined analogous to Lustre's \emph{sequence operators} and illustrated in Figure~\ref{fig:memory-operators}. Note that special attention is required when combining memory with other side effects. The evaluation of \textbf{pre} $e$ exhibits the side-effects of $e$ already in the first evaluation cycle. Similarly, $e_1\, \bm{\rightarrow}\, e_2$ exhibits the side-effects of both sub-expressions in the first cycle as well. In $e_1$ \textbf{fby} $e_2$, the side-effects of $e_2$ are only observed from the second cycle on. 

Extensions with respect to Lustre are the two \emph{optional value constructors} \textbf{Some} and \textbf{None}, and their \emph{match} construct (\textbf{either}/\textbf{or}). These are important for defining optional inputs and outputs, as seen in the example in Section~\ref{sec:mimosa-lang}. For an expression $\textbf{either}\, e_1\, \textbf{or}\, e_2$, first $e_1$ is evaluated. If $e_1$ evaluates to $\textbf{Some}\; t$, the overall value of the match construct is $t$. If $e_1$ evaluates to \textbf{None}, then $e_2$ is evaluated, which becomes the overall value of the match construct. Note that the second expression is evaluated only on demand, so that possible side-effects of $e_2$ are only observed when $e_1$ is \textbf{None}.

Mimosa employs a Hindley-Milner~\cite{damas-type-schemes} style type system, whose formal description we omit for brevity. We assume that the base types \texttt{unit}, \texttt{bool}, \texttt{int}, and \texttt{float} are present, together with their respective constants and usual arithmetic and logic functions (implicitly defined as external steps).

\begin{figure}
    \centering
    \includegraphics[width=0.6\linewidth]{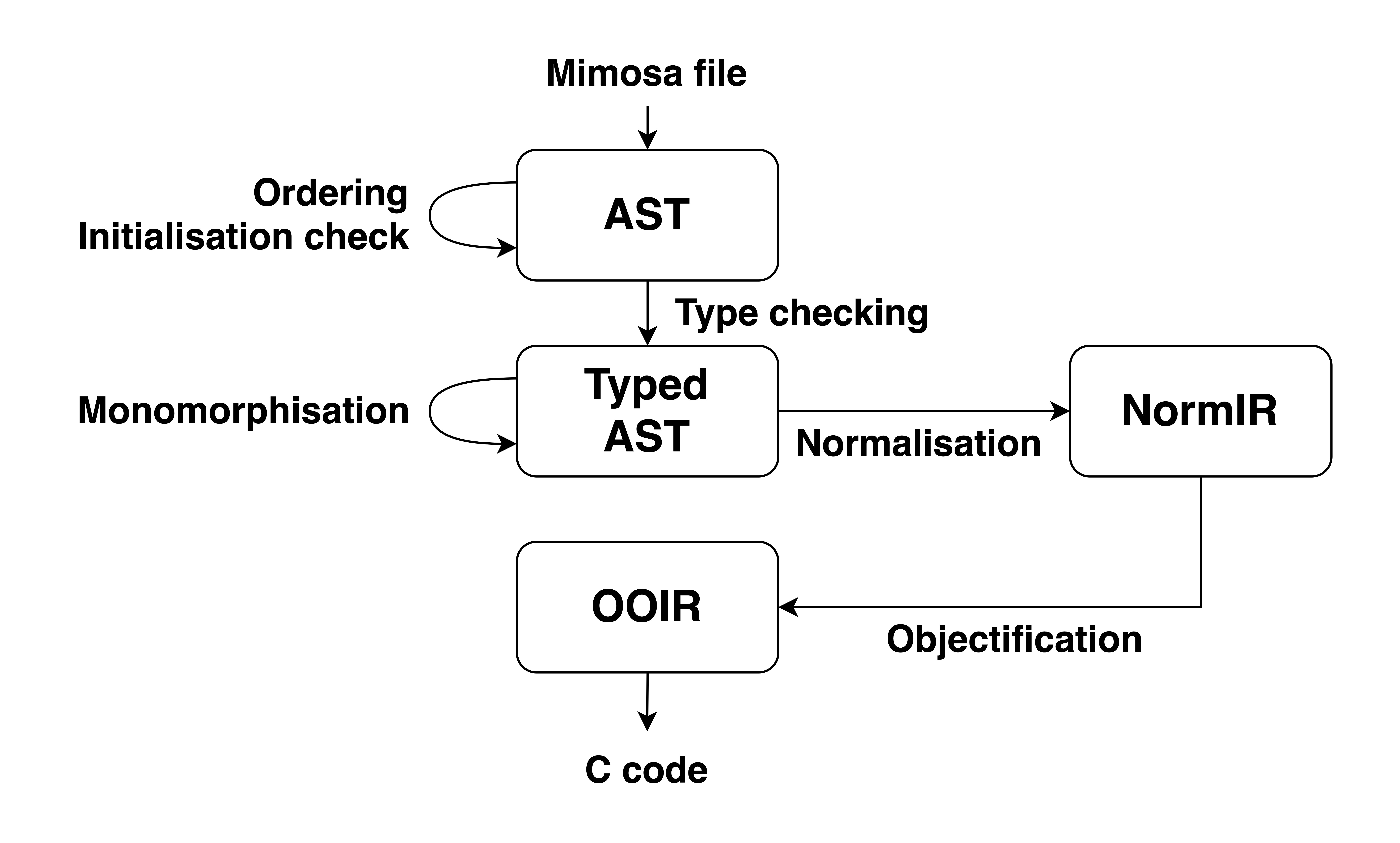}
    \caption{Compiler phases overview}
    \label{fig:compiler-overview}
\end{figure}

Figure~\ref{fig:compiler-overview} illustrates the different phases of step compilation. A Mimosa file is parsed into an abstract syntax tree (AST), on which multiple analyses are performed. Similar to Lustre, the order of equations inside a step is not significant, as the compiler performs a \emph{dependency analysis} and brings them into a suitable order. This also rejects programs which have any non-causal cyclic dependencies. The \emph{initialisation check}~\cite{colaco-lustre-init} tries to prove that the undefined value $\bot$ introduced by \textbf{pre} does not influence the result of the overall step evaluation. 

Steps in Mimosa can be \emph{polymorphic}. For example, a generic identity function may be implemented as:

\begin{lstlisting}[language=mimosa, basicstyle=\footnotesize]
    step id (a : 'a) --> (b : 'a) { b = a; }
\end{lstlisting}

Polymorphic steps are compiled away during \emph{monomorphisation}, which first records all call-sites of a step, and then creates copies for each required type instantiation. This is similar to how polymorphic functions are compiled in other languages, such as Rust~\cite{klabnik2019rust} and SML~\cite{sml-mono}, and how low-level code is extracted from Coq~\cite{akira-coq-mono}.

For brevity, we only describe the \emph{normalisation} and \emph{objectification} phases in detail, after which the extraction of sequential code is trivial. The overall compilation scheme is similar to the one presented for Lustre~\cite{biernacki-clock-directed-compilation}, with adaptations for the specific evaluation strategy of Mimosa.

\begin{figure}[t]
{\small
\begin{align*}
\begin{array}{r@{\quad}c@{\quad}l@{\quad}l}
b : \text{Base} & ::= & x & \text{Variable} \\
{} & | & c & \text{Constant} \\
{} & | & \textbf{None} & \text{None} \\
{} & | & \textbf{Some}\, x & \text{Some} \\
e : \text{Expression} & ::= & b & \text{Base expression} \\
{} & | & x\bm{,}\, \ldots\bm{,}\, x & \text{Tuple} \\
{} & | & \textbf{pre}\, x & \text{Pre} \\
{} & | & bl\, \textbf{fby}\, bl & \text{Followed-by} \\
{} & | & x\, x & \text{Application} \\
{} & | & \textbf{if}\, x\, \textbf{then}\, bl\, \textbf{else}\, bl & \text{Conditional} \\
{} & | & \textbf{either}\, x\, \textbf{or}\, bl & \text{Option match} \\
bl : \text{Block} & ::= & [\, p_i\, =\, e_i\, ]^n,\, b & \text{Block}
\end{array}
\end{align*}}
\caption{NormIR}
\label{fig:norm-ir}
\end{figure}

\textbf{Normalisation} The normalisation pass aims to transform the typed AST into a \emph{normalised intermediate representation} (NormIR). The abstract syntax of NormIR is shown in Figure~\ref{fig:norm-ir}. It differs from the typed AST in various ways. Expressions are divided into \emph{base expressions} (designating base values such as variables and constants), and standard \emph{expressions}. Nesting is made explicit through \emph{block structure}. A block consists of a list of \emph{normalised equations}, where the pattern is now either a simple variable or a list of variables (without nesting), and a base expression. Conceptually, one can think of a block in terms of nested \textbf{let/in} expressions in a functional language like OCaml or Haskell:

{\footnotesize \begin{align*}
\left [
\begin{array}{c@{\quad}c@{\quad}c}
x & = & 2 \\
y & = & x + 4
\end{array}
\right ], y
\end{align*}}

\noindent The above block is essentially the same as the following expression in OCaml:

\begin{center}
\begin{tabular}{c}
\begin{lstlisting}[language=caml, basicstyle=\footnotesize]
let x = 2 in
let y = x + 4 in
y
\end{lstlisting}
\end{tabular}
\end{center}

In NormIR, sub-expressions are either a \emph{block}, which provides a new scope for nesting, or a \emph{named reference} (i.e. a variable). For simplicity, we assume that the patterns in equations are not nested. Otherwise, they can always be rewritten:

{\footnotesize \begin{gather*}
x, (y, z) = e; \\
\Downarrow \\
x, \langle tmp \rangle = e; \\
y, z = \langle tmp \rangle;
\end{gather*}}

In all translation rules in this paper, we adopt the notation $\langle tmp \rangle$ to denote a fresh identifier, i.e. a variable name not yet bound in the current scope.

\begin{sidewaysfigure}[p]
\centering
\begin{gather*}%
    \begin{aligned}
    % Variables
    \frac%
    { }%
    {\comp{x}{[\;]}{x}}% 
    \; \texttt{\small (Var)}%
    & \; &
    % Constants
    & \frac%
    { }%
    {\comp{c}{[\;]}{c}}%
    \; \texttt{(Const)}%
    & \; &
    % Some
    \frac%
    { e\, \Rrightarrow\, s,\, e' }%
    { \textbf{Some}\, e\, \Rrightarrow\, (s\, ::\, [\, \langle r \rangle\, =\, \textbf{Some}\, e'\,]), \, \langle r \rangle}%
    \; \texttt{(Some)} %
    \end{aligned}\\[4mm]%
    \begin{aligned}
    % None
    \frac%
    { }%
    { \textbf{None}\, \Rrightarrow\, [\;],\, \textbf{None} }%
    \; \texttt{(None)}
    & \; &
    % Init
    \frac%
    {e_1\, \Rrightarrow b \qquad \comp{e_2}{s_2}{e_2'}}%
    {\comp{e_1\, \rightarrow\, e_2}{(s_2\, :: [\, \langle y \rangle\, =\, e_2';\, \langle r \rangle\, =\, b\, \textbf{fby}\, ([\;],\, \langle y \rangle)\,])}{\langle r\rangle}}%
    \; \texttt{(Init)}
    \end{aligned}\\[4mm]%
    \begin{aligned}
    % Pre
    \frac%
    {\comp{e}{s}{e'}}%
    {\textbf{pre}\,e\, \Rrightarrow\, (s \, ::\, [\, \langle x \rangle \,=\, e';\, \langle r \rangle \,=\, \textbf{pre}\, \langle x \rangle \,]),\, \langle r \rangle}%
    \; \texttt{(Pre)}%
    & \quad &
    % Fby
    \frac%
    {e_1\, \Rrightarrow\, b_1 \qquad e_2\, \Rrightarrow\, b_2}%
    {\comp{e_1\, \textbf{fby}\, e_2}{[\,\langle r \rangle\, =\, b_1\, \textbf{fby}\, b_2\,]}{\langle r \rangle}}%
    \; \texttt{(Fby)}%
    \end{aligned} \\[4mm]%
    % Tuples
    \frac%
    {\comp{e_1}{s_1}{e_1'} \qquad \ldots \qquad \comp{e_n}{s_n}{e_n'}}%
    {e_1,\, \ldots,\, e_n\, \Rrightarrow\, (s_1\, :: \ldots\, ::\, s_n\, ::
    [\,\langle t_1 \rangle\, =\, e_1';\, \ldots;\, \langle t_n \rangle\, =\, e_n';\, \langle r \rangle\, =\, \langle t_1 \rangle,\, \ldots,\, \langle t_n \rangle\,]),\, \langle r \rangle}
    \; \texttt{(Tuple)} \\[4mm]%
    % If-then-else
    \frac%
    {\comp{e_1}{s}{e_1'} \qquad e_2\, \Rrightarrow\, b_2 \qquad e_3\, \Rrightarrow\, b_3}%
    {\textbf{if}\, e_1\, \textbf{then}\, e_2 \, \textbf{else}\, e_3\, \Rrightarrow (s\, ::\, [\, \langle x \rangle\, =\, e_1';\, \langle r\rangle\, =\, \textbf{if}\, \langle x \rangle\, \textbf{then}\, b_2\, \textbf{else}\, b_3\,],\,\langle r \rangle
    }%
    \; \texttt{(If)} \\[4mm]%
    \begin{aligned}
    % Either-or
    \frac%
    {e_1\, \Rrightarrow\, s,\, e_1' \qquad e_2\, \Rrightarrow\, b}%
    {\textbf{either}\, e_1\, \textbf{or}\, e_2\, \Rrightarrow (s\, ::\, [ \, \langle x \rangle\, =\, e_1';\, \langle r \rangle\, =\, \textbf{either}\, \langle x \rangle\, \textbf{or}\, b\, ]),\, \langle r \rangle}%
    \; \texttt{(Either)} %
    & \; & 
    % Application
    \frac%
    {e_1\, =\, f \qquad e_2\, \Rrightarrow\, s,\, e_2'}%
    {e_1 \; e_2 \, \Rrightarrow\, (s\, ::\, [\, \langle x \rangle\, =\, e_2';\, \langle r \rangle\, =\, f\; \langle x \rangle\,]),\, \langle r \rangle}%
    \; \texttt{(App)}
    \end{aligned}%
\end{gather*}
\vspace{0.2cm}
\caption{Normalisation of expressions}
\label{fig:expression-normalisation}
\end{sidewaysfigure}

The general translation rules shown in Figure~\ref{fig:expression-normalisation} describe how to transform each expression into a block.

This is trivial for variables and constants, as well as the \textbf{None} constructor. For $\textbf{Some}\, e$, the expression $e$ is lifted to the surrounding scope and its value bound to a new name, which can be used as the variable name for the NormIR \textbf{Some} constructor.

The memory operators are more interesting, as NormIR only retains the \textbf{fby} and \textbf{pre} forms. In Mimosa, \textbf{pre} $e$ cannot simply be rewritten in terms of \textbf{fby} as is done in the Lustre compiler, since any potential side-effect of the evaluation of $e$ predates the value of $e$ by one evaluation cycle. We therefore keep \textbf{pre} as its own operator in NormIR. The only thing to do is to move the evaluation of $e$ to the surrounding scope, thereby forcing its evaluation eagerly. The expression $e_1\, \textbf{fby}\, e_2$ is the easiest to translate, as the resulting blocks $b_1$ and $b_2$ from translating $e_1$ and $e_2$, respectively, are just used as the arguments of the \textbf{fby} operator in NormIR. The initialisation expression $e_1 \bm{\rightarrow} e_2$ is translated by moving $e_2$ to the surrounding scope (thereby forcing its evaluation eagerly again).

Tuples are translated by separately transforming each element to a block, and assigning a new name $\langle t_i \rangle$ to each resulting base expression. The equations of the final block are then the concatenation of all the equations resulting from this transformation. The final equation is a tuple construction only using the newly assigned names $\langle t_i \rangle$.

The remaining cases follow similarly, relying on the trick of hoisting sub-expressions that always need to be evaluated (e.g., the condition $e_1$ in \texttt{If} or $e_1$ in \texttt{Either}) to the surrounding scope to force their evaluation, while leaving sub-expressions which shall only be evaluated when needed (e.g., the two branches $e_2$ and $e_3$ in \texttt{If} or $e_2$ in \texttt{Either}) in their own block structure.

\begin{figure}
\footnotesize
\begin{gather*}
\textbf{step}\, name\, \bm{(}\, i_1\bm{,}\, \ldots\bm{,}\, i_l\, \bm{)}\, \bm{\longrightarrow}\, \bm{(}\, o_1\bm{,}\, \ldots\bm{,}\, o_k\, \bm{)}\, [\, p_i\, =\, e_i\, ]^n\\
\Downarrow \\
e_1\, \Rrightarrow\, s_1,\, e_1' \qquad \ldots \qquad e_n\, \Rrightarrow\, s_n,\, e_n' \\
\Downarrow \\
\textbf{step}\, name\, \langle in \rangle\, \bm{\longrightarrow}\, \langle out \rangle\, (
\left (
\begin{array}{c}
    [\, i_1\bm{,}\, \ldots\bm{,}\, i_l\, =\, \langle in \rangle\,]\\
    \vdots \\
    ::\, s_i\, ::\, [\, p_i\, =\, e_i'\, ]\, :: \\
    \vdots \\
    \left [\, \langle out \rangle\, =\, o_1\bm{,}\, \ldots\bm{,}\, o_k\, \right ]
\end{array}
\right )
, \langle out \rangle )
\end{gather*}
\caption{Normalisation of steps}
\label{fig:step-normalisation}
\end{figure}

Finally, each step is normalised by creating new input and output variables $\langle in \rangle$ and $\langle out \rangle$ and normalising each equation as shown in Figure~\ref{fig:step-normalisation}. This step is necessary for extracting sequential code, which most often allows functions to have only a single return value.

\begin{figure}
{\small
\begin{align*}
\begin{array}{r@{\quad}c@{\quad}l@{\quad}l}
e : \text{Expression} & ::= & x & \text{Variable} \\
{} & | & \textbf{!}\, x & \text{State variable} \\
{} & | & c & \text{Constant} \\
{} & | & \textbf{None} & \text{None} \\
{} & | & \textbf{Some}\, x & \text{Some} \\
i : \text{Instruction} & ::= & x\, =\, e & \text{Assignment} \\
{} & | & x\, \bm{\leftarrow}\, e & \text{State assignment} \\
{} & | & x\, =\, x\bm{,}\, \ldots\bm{,}\, x & \text{Tuple construction} \\
{} & | & x\bm{,}\, \ldots\bm{,}\, x\, =\, x & \text{Tuple destruction} \\
{} & | & x.\textbf{reset} (\, x\, ) & \text{State reset} \\
{} & | & \textbf{return}\, e & \text{Return} \\
{} & | & \makecell[l]{\textbf{if}\, x\, \textbf{then}\, [\, i\,]\,  \textbf{else}\, [\, i\, ]} & \text{Conditional} \\
{} & | & x\, =\, x.\textbf{step}\, \bm{(}\, x,\, x\, \bm{)} & \text{Application} \\
{} & | & \makecell[l]{\textbf{case}\, x\, \bm{\ \{} \, \textbf{Some}\, x:\, [\, i \,]; \\ \textbf{None}: \, [\, i \,]\, \textbf{\}}} & \text{Option match} \\
d : \text{Machine} & ::= & (\, m,\, j,\, r,\, s\, ) & {} \\
m,\, j & ::= & [\, x \,] & {} \\
r,\, s & ::= & [\, i\, ]
\end{array}
\end{align*}}
\caption{Object-oriented IR}
\label{fig:ooir}
\end{figure}

\textbf{Objectification} After normalisation, scope is explicit, and the different memory operators are expressed in only two forms. Using equations, the program's overall structure remains declarative. The next phase of the compiler, \emph{objectification}, which makes state explicit, introduces a shift from a declarative to an imperative paradigm. As state is a well-known concept in object-oriented programming languages, we adopt an \emph{Object-Oriented Intermediate Representation} (OOIR) as the output of this phase. The overall structure of OOIR is shown in Figure~\ref{fig:ooir}. 

Each step is translated into a \emph{machine} (following the nomenclature used by the Lustre compiler~\cite{biernacki-clock-directed-compilation}). Each machine consists of four parts: a \emph{memory} $m$, a set of \emph{instance variables} $j$, a set of \emph{reset instructions} $r$, and a set of \emph{step instructions} $s$. The memory $m$ contains all the variables that need to be remembered across step evaluations, and the instance variables $j$ refer to the instantiated machines used within this machine (as each application of a step within another step needs to keep its own state). The reset instructions $r$ initialise the memory $m$ and the instantiated machines in $j$, while the step instructions $s$ perform the actual computation of the step.

Expressions in OOIR distinguish between variable and state references (which are variables that are remembered beyond a single evaluation of a step). After objectification, the computation performed by each step is described as a list of \emph{instructions}. The most straightforward instructions are \emph{variable} and \emph{state assignment}. \emph{Tuple construction} and \emph{destruction} only refer to simple names, and there is no nesting. A \emph{state reset} $t.\textbf{reset} (o)$ runs the reset instructions of machine $t$ on the particular instance $o$. Conceptually, \textbf{reset} acts similarly to a constructor in object-oriented programming languages, where $t$ would represent the class, and $o$ a particular object instance. The \emph{return} and \emph{conditional} instructions are trivial. The \emph{step application} $r = t.\textbf{step} (a, o)$ runs the step instructions of machine $t$ on the particular instance $o$ with argument $a$, and binds the returned value to $r$. The \emph{option match} instruction performs a case distinction for the two optional type constructors.

The overall translation from NormIR to OOIR is shown in Figure~\ref{fig:normir-to-ooir}. It uses two different relations: $m \triangleright e \rightsquigarrow e'$ expresses that under a memory $m$, the base expression $e$ in NormIR translates to an expression $e'$ in OOIR. These rules are straightforward for most cases. To translate a variable reference, we need to check whether the particular variable is part of the memory of a step. If so, the resulting OOIR expression is a \emph{state variable}.

The relation $d \triangleright eq \Rightarrow d'$ expresses that a machine $d$ is transformed into a machine $d'$ by translating the equation $eq$. By slight abuse of notation, we use the same relation also for lists of equations, in which case the translation should be understood as a left-fold over the list:
{\begin{align*}
\frac
{}
{d \triangleright [\;] \Rightarrow d}
&
\quad
&
\frac
{d \triangleright eq \Rightarrow d' \quad d' \triangleright eqs \Rightarrow d''} 
{d \triangleright eq :: eqs \Rightarrow d''}
\end{align*}}

\vspace{2mm}

We illustrate the translation with an example. From the \texttt{edge} step in Figure~\ref{fig:edge-detector-mimosa} we have the following equation:

\begin{lstlisting}[language=mimosa, basicstyle=\footnotesize]
  pre_in = in -> pre in;
\end{lstlisting}

\noindent Normalisation, according to the \texttt{Init} and \texttt{Pre} rules in Figure~\ref{fig:expression-normalisation}, results in the following equations (after inlining of equivalent variable names):

\begin{lstlisting}[language=mimosa, basicstyle=\footnotesize, morekeywords={if,then,else,true,false, fby}]
  tmp0 = pre in;
  pre_in = in fby tmp0;
\end{lstlisting}

\noindent This essentially forces the evaluation of the right-hand side of the $\bm{\rightarrow}$ operator. The translation to OOIR involves the creation of \emph{state variables}. For the \textbf{pre} statement, a state variable \texttt{tmp1} is introduced, which is initialised to a dummy value \texttt{nil} in the reset instructions of the resulting machine. In practice, this value is chosen as an arbitrary constant according to the type of the variable \texttt{in}, as the initialisation check has already proven that the initial value of any \textbf{pre} statement does not alter the overall step result. In the step instructions, the value of \texttt{tmp0} is then the current value of \texttt{tmp1}, which gets updated with the current value of \texttt{in} afterwards.

The basic principle of compiling \textbf{fby} is to define a state variable \texttt{first}, which records if this is the first time the expression is evaluated, and using its value as the condition for an imperative conditional statement. Together, this results in the following code:

\begin{lstlisting}[escapeinside={(§}{§)},morekeywords={if,then,else,true,false}, basicstyle=\footnotesize]
  (* part of the reset instructions *)
  tmp1 (§$\bm{\leftarrow}$§) nil
  first (§$\bm{\leftarrow}$§) true

  (* part of the step instructions *)
  tmp0 = (§\textbf{!}§)tmp1
  tmp1 (§$\bm{\leftarrow}$§) in
  if (§\textbf{!}§)first then
    [ pre_in = in ]
  else
    [ pre_in = tmp0 ]
  first (§$\bm{\leftarrow}$§) false
\end{lstlisting}

Each step body (after normalisation) is described by a single block $({eqs},\, e)$. To translate a step to OOIR, the list of equations ${eqs}$ is translated starting from an empty machine
$([\,], [\,], [\,], [\,]) \triangleright {eqs} \Rightarrow (m, j, r, s)$, and the base expression $e$ is translated $m \triangleright e \rightsquigarrow e'$. The overall OOIR representation of a step is then $(m, j, r, s :: \, [\, \textbf{return}\,  e'\,])$.

The compilation of steps leads to a set of machines, from which sequential code extraction is straightforward. In the case of C, each machine results in two functions (the reset and step functions) and a structure defining the state. We can therefore assume that the steps presented in the example in Figure~\ref{fig:edge-detector-mimosa} result in C code providing the following signatures:

\begin{lstlisting}[
    language=C, 
    basicstyle=\footnotesize
]
  struct edge_state_t
  void edge_reset (struct edge_state_t *)
  struct opt_bool edge_step (bool, struct edge_state_t *)

  struct toggle_state_t
  void toggle_reset (struct toggle_state_t *)
  void toggle_step (bool, struct toggle_reset_t *)
\end{lstlisting}

Prototypes are only included with their signature. Their implementation must be provided by external C code during compilation. The step compiler is also responsible for creating the type definitions for optional types and tuples, such as the type \textbf{struct} \texttt{opt\textunderscore bool}, which is defined as

\begin{lstlisting}[language=C, basicstyle=\footnotesize]
  struct opt_bool
  {
    bool is_some;
    bool value;
  }
\end{lstlisting}

\begin{sidewaysfigure}[p]
\centering
\begin{gather*}
\begin{aligned}
\frac{ }{ m \triangleright c \rightsquigarrow c }\; \texttt{(Const)} & \quad &
\frac{ v \notin m }{ m \triangleright v \rightsquigarrow v}\; \texttt{(Var)} & \quad &
\frac{ v \in m}{ m \triangleright v \rightsquigarrow \textbf{!}\, v}\; \texttt{(StateVar)} & \quad &
\frac{ }{m \triangleright \textbf{None} \rightsquigarrow \textbf{None}}\; \texttt{(None)} \end{aligned}\\[2mm]
\begin{aligned}
\frac{ }{ m \triangleright \textbf{Some}\, v \rightsquigarrow \textbf{Some}\,v} \; \texttt{(Some)} & \quad & 
\frac%
{ m \triangleright e \rightsquigarrow e' }%
{ (m, j, r, s) \triangleright v = e \Rightarrow (m, j, r, (s ::\, [\, v = e'\,]))}%
\; \texttt{(Base)}
\end{aligned} \\[2mm]
\frac{(m, j, r, [\;]) \triangleright s_t \Rightarrow (m', j', r', s_t') \qquad m' \triangleright e_t \rightsquigarrow e_t' \qquad (m', j', r', [\;]) \triangleright s_e \Rightarrow (m'', j'', r'', s_e') \qquad m'' \triangleright e_e \rightsquigarrow e_e'}%
{ (m,j,r,s) \triangleright x = \textbf{if}\, c\, \textbf{then}\, (s_t, e_t)\, \textbf{else}\, (s_e, e_e) \Rightarrow (m'', j'', r'', (s\, :: \, [\, \textbf{if}\, c\, \textbf{then}\, (s_t'\, ::\, [\, x\, =\, e_t'\,])\, \textbf{else}\, (s_e'\, ::\, [\, x\, =\, e_e'\,])\,])) }%
\; \texttt{(If)} \\[3mm]
\frac%
{}%
{(m, j, r, s) \triangleright v = \textbf{pre}\, x \Rightarrow ((m\, ::\, [\, \langle tmp \rangle\,]), j, (r\, ::\, [\,\langle tmp \rangle \bm{\leftarrow} nil\,]), (s \, ::\, [\, v = \textbf{!}\, \langle tmp \rangle;\, \langle tmp \rangle \bm{\leftarrow} x \,]))}
\; \texttt{(Pre)} \\[2mm]
\frac%
{(m, j, r, [\;]) \triangleright s_1 \Rightarrow (m', j', r', s_1') \qquad m' \triangleright e_1 \rightsquigarrow e_1' \qquad (m', j', r', [\;]) \triangleright s_2 \Rightarrow (m'', j'', r'', s_2') \qquad m'' \triangleright e_2 \rightsquigarrow e_2'}%
{\makecell[c]{(m, j, r, s) \triangleright v = (s_1, e_1) \, \textbf{fby}\, (s_2, e_2) \Rightarrow \\ ((m''\, ::\, [\,\langle \mathit{fst} \rangle\,]), j'', (r''\, :: \, [\, \langle \mathit{fst} \rangle \bm{\leftarrow} true \,]), (s\, ::\, [\, \langle t \rangle = \textbf{!}\, \langle \mathit{fst} \rangle;\, \textbf{if}\, \langle t \rangle\, \textbf{then}\, (s'\, ::\, [\, v = e_1'\,])\, \textbf{else}\, (s''\, ::\, [\, v = e_2'\,]);\, \langle \mathit{fst} \rangle \bm{\leftarrow} false\,]))}}
\; \texttt{(Fby)} \\[2mm]
\frac%
{(m,j,r,[\;]) \triangleright s \Rightarrow (m', j', r', s') \qquad m' \triangleright e \rightsquigarrow e'}%
{(m,j,r,s) \triangleright v = \textbf{either}\, x\, \textbf{or}\, (s, e) \Rightarrow (m', j', r', (s\, ::\, [\, \textbf{case}\, x\, \bm{\{} \textbf{Some}\, y: [\, v = y\, ]; \textbf{None}: (s'\, ::\, [\, v = e'\,] \bm{\}}]))}%
\; \texttt{(Either)} \\[2mm]
\frac%
{ }%
{ (m,j,r,s) \triangleright v = \mathrm{f}\, x \Rightarrow (m, (j\, ::\, [\, \langle o \rangle\,]), (r\, ::\, [\, \mathrm{f}.\textbf{reset} (\, \langle o \rangle\, )\,]), (s\, ::\, [ \, x = \mathrm{f}.\textbf{step} (\, x,\, \langle o \rangle\, )\,]))}
\; \texttt{(App)} \\[2mm]
\frac{ }{(m, j, r, s) \triangleright v = x_1, \ldots, x_n \Rightarrow (m, j, r, (s\, ::\, [\, v = x_1, \ldots, x_n\,]))}\; \texttt{(TupleConstr)} \\[2mm]
\frac%
{ (m, j, r, s) \triangleright \langle t \rangle = e \Rightarrow (m', j', r', s') }%
{ (m,j,r,s) \triangleright v_1\bm{,}\, \ldots\bm{,}\, v_n = e \Rightarrow (m',j',r', (s'\, ::\, [\, v_1\bm{,}\, \ldots\bm{,}\, v_n = \langle t \rangle\, ]))}%
\; \texttt{(TupleDestr)}
\end{gather*}
\vspace{0.2cm}
\caption{Translation from NormIR to OOIR}
\label{fig:normir-to-ooir}
\end{sidewaysfigure}

\subsection{Compiling channels and nodes}
\label{sec:compile-coordination-layer}

The coordination layer of Mimosa (including the channels and nodes) needs to be translated into primitives provided by an RTOS. The translation shown here uses generic types and API names which different RTOSs can provide.

Each channel is translated to a FIFO queue. In theory, the FIFOs could be implemented as a linked list to grow and shrink automatically as needed. The queue primitives provided by most RTOSs utilise a more compact and performant implementation backed by an array, which means that the upper bound on the size of each channel must be known at compile time. This is done by providing a \emph{model} to the compiler, which lists the size of each buffer defined in the Mimosa input. We explain this model file in more detail below.

The stringent timing requirements of MIMOS (reading at the beginning and writing at the end of a period) cannot, in general, be met by software that implements the nodes as tasks or processes. For example, on a sequential processor, it is impossible to write multiple items simultaneously to different FIFOs. The beginning of a period can also only be interpreted as the release of the task, not the exact time point at which the computation is performed.

Therefore, the timing semantics of Mimosa has to be faithfully emulated. To do so, each item in a queue carries a time stamp, which marks the time point at which it is supposed to be written according to the semantics (i.e., becomes available for consumption). When writing outputs, each node assigns a calculated time stamp to each written data item, which corresponds to the end of its current period. A task reading from a channel can then check these time stamps against its current release time to determine whether an item is available for consumption.  In this way, the required timing semantics can be emulated as long as the worst-case response time of each node is shorter than its period.

For example, channel \texttt{a} from Figure~\ref{fig:edge-detector-mimosa} is translated to the following code:

\begin{lstlisting}[language=C, basicstyle=\footnotesize]
    queue_t a = 
      create_queue (A_SIZE, sizeof (bool))
    queue_t a_stamps = 
      create_queue (A_SIZE, sizeof (timestamp_t))
\end{lstlisting}

\noindent We use a second queue to store the time stamps. Equivalently, each data item could be wrapped in a structure that holds both the data and the time stamp.

The parameter \texttt{A\textunderscore SIZE} is the bound on the size of the channel, and must, as described before, be provided as a model to the compiler. Additionally, the model must provide the priorities and stack sizes for all generated tasks (i.e., for all nodes of the Mimosa program). This means that schedulability analysis has to be performed externally, and automated estimation of channel bounds is part of our ongoing research work~\cite{wang-mimos-tool}.

The task function implementing the \texttt{edge} node has an overall structure as presented in Figure~\ref{fig:example-task-c}. After setting up the state of the implemented step, the task enters an infinite loop, where it first checks the availability of data in the input buffer (by comparing the time stamp against the current time). If data is available, the input data item is retrieved from the input queue (including the removal of the corresponding time stamp), and the step function is invoked. As the \texttt{edge} node has an optional output, the return value of the step is unwrapped (i.e. the \texttt{is\textunderscore some} field is checked, and if it is true, the respective value is written to the output queue). Additionally, the time stamp of the end of the current node activation (i.e. the next activation time) is written to the corresponding time stamp queue of the output channel. Afterwards, the task sleeps until the next activation time.

\begin{figure}
\begin{lstlisting}[
    language=C, 
    basicstyle=\footnotesize,
    numbers=left,
    numbersep=0pt,
    numberstyle=\tiny
]
  void edge () {
    /* Create/reset step state */
    struct edge_state_t self;
    edge_reset(&self);

    time_t now = 0;

    while(1) {
      /* Calculate end of period */
      time_t next_period = now + edge_period;
      
      /* Check if input data is available */
      bool a_avail = check_avail(now, a_stamps);
      if (a_avail) {
        /* Receive input data item */
        bool a_val = queue_recv(a);
        
        /* Remove corresponding time stamp */
        queue_recv(a_stamps);
        
        /* Call implemented step function */
        struct opt_bool r = 
            edge_step (a_val, &self);
            
        /* Unwrap optional output */
        if (r.is_some) {
            /* Write at end of current period */
            queue_send(b, r.value);
            queue_send(b_stamps, next_period);
        } 
      }
      now = next_period;
      
      /* Wait till next period */
      sleep_until (next_period);
    }
  }
\end{lstlisting}
\caption{Example of task function in C}
\label{fig:example-task-c}
\end{figure}

Note that some RTOSs (e.g., FreeRTOS\footnote{https://www.freertos.org}) provide specific \texttt{delay\textunderscore until} functions which prevent the release time from drifting due to the execution time of the task function. The exact setup depends on the particular RTOS.
\section{Related work}
\label{sec:related-work}

The compilation scheme for Lustre as presented in~\cite{biernacki-clock-directed-compilation} represents an advancement over the original way of compiling~\cite{caspi-lustre}, which relied on inlining node definitions until only a single node remained, from which it is easy to generate sequential code and a state vector. The clock-directed, modular translation scheme presented in~\cite{biernacki-clock-directed-compilation} has the advantage that each node can be implemented as a separate function, which holds the potential for significant decreases in code size.

The idea of utilising an RTOS as the underlying execution environment for a programming language offering multiple execution threads is not new. 
For example, the programming language PEARL~\cite{brandes-pearl} enables the definition of tasks and the management of coordination between them through various communication primitives, making it suitable for real-time applications. Currently, the most well-known and widely used programming language in this category is Ada. The Ravenscar profile~\cite{burns-ravenscar} describes a runtime profile for Ada, which explicitly targets embedded applications, and can either be run on top of a chosen RTOS, or on a minimal runtime, which usually comes with the respective Ada compiler~\cite{ravenscar-gnat}.

Mimosa is neither the first nor the only programming language designed for real-time embedded software applications. In the following, we shortly present some of the more recent research efforts in this area, of which we are aware:

The Timed-C~\cite{natarajan-timed-c} language extends C with timing primitives. It defines tasks that can communicate through channels, similar to those in MIMOS. The Timed-C compiler~\cite{timed-c-compiler} is a source-to-source compiler, which compiles to plain C on top of either the real-time POSIX profile or FreeRTOS.

The family of synchronous languages, including Lustre~\cite{caspi-lustre} (and its commercial counterpart SCADE~\cite{colaco-scade}), Esterel~\cite{berry-esterel}, and Signal~\cite{benveniste-signal} bring the same synchronous abstraction into software as those used for digital circuit design. By utilising a global system tick, which is taken as the smallest observable time interval, a programmer can abstract over the individual execution times of the various software components constituting the overall system. This shift from physical real-time to logical time ticks simplifies temporal reasoning about the behaviour of a program. It allows verifying correctness independently of the timing characteristics of the actual execution platform. The stringent timing requirements of the synchronous execution model are challenging to meet on modern hardware, which often incorporates many and multi-core processor units, as well as distributed execution platforms. MIMOS, and by extension Mimosa, attempts to circumvent some of these problems by adopting an asynchronous execution model. The compilation scheme presented in this paper only deals with compiling to tasks, but is agnostic to the particular scheduling algorithm used. It would be a straightforward extension to distribute the set of tasks over multiple processing cores.

Giotto~\cite{henzinger-giotto} is a programming language specifically designed for real-time systems. Similar to Mimosa, a program is composed of periodic tasks that communicate through ports. Each port always keeps the latest value that was written to it. One particular focus of Giotto lies in execution modes. A microkernel~\cite{kirsch-giotto-microkernel} has been introduced as a compilation target, which separates the scheduling of the various tasks from the interaction with the environment through a dedicated virtual machine called E-Machine~\cite{henzinger-e-machine}. A similar approach might also be an interesting alternative for implementing Mimosa.

Lingua Franca (LF)~\cite{lohstroh-lingua-franka} offers a polyglot framework for expressing concurrency and coordination in various programming languages, with a focus on hard real-time systems. It is built upon the Reactor model~\cite{lohstroh-reactors}, where computation is triggered by different events. LF augments other programming languages with a coordination layer that has discrete event semantics, which means that the LF runtime must be implemented for each programming language it supports. Deantoni et al.~\cite{lf-gemoc} incorporated LF's coordination semantics into the Eclipse-based language and modelling workbench Gemoc~\cite{bousse-gemoc}. Through this, they created an \emph{omniscient} interactive debugger (i.e. one in which the user can step forward and backwards in time), as well as various other tooling, such as graphical debugging and modelling facilities. A similar approach might also offer interesting opportunities for future work on Mimosa. 
\section{Conclusion and Future Work}
\label{sec:conclusion}

This paper presents a compilation scheme for programs written in the Mimosa programming language. This is accomplished by separately compiling the computation steps into sequential code and translating the coordination layer, which consists of nodes and channels, into primitives provided by an RTOS.

Mimosa is still a prototype language, and many additional features need to be added to make it useful for implementing complex embedded system software.

The reading and writing policies for each port of a node can be relaxed, allowing multiple data items to be read or written each time a node is activated. This can be even further relaxed to allow reading or writing \emph{up-to k} elements at any particular port~\cite{wang-mimos-tool}. These extensions enable nodes with different periods to communicate without encountering overflow or underflow issues in the channel between them. Mimosa's expression language could be extended with operators on lists of values, such as those found in functional programming languages, including \emph{map} and \emph{fold}. This is similar to the extension of Lustre with array iterators~\cite{morel-lustre-array-iterators}.

Similarly, \emph{registers}~\cite{wang-mimos-tool} can be introduced as a communication link, which always retain the latest value written to them. As it has been shown in the original MIMOS paper~\cite{wang-mimos}, these types of extensions to the reading and writing policies of a node can be done without affecting determinism, as the reading and writing times are fixed by the period and deadline of each node.

The shift from a strict evaluation of sub-expressions (for example, in the conditional expression) in Lustre to a non-strict evaluation in Mimosa opens the possibility of including side-effectful expression evaluation in the language. This is illustrated in the example in Figure~\ref{fig:edge-detector-mimosa}, where the function toggling the LED is only selectively called depending on the input value of the step. We have chosen this design to offer a mechanism similar to Lustre's multi-clocks, but without the mental overhead of manually defining multiple clocks. In a future version of Mimosa, we would like to incorporate the possibility of tracking effects in the type system, as is done in various recent programming languages (e.g., Flix~\cite{madsen-flix-lang}, Koka~\cite{reinking-perceus}, Effekt~\cite{brachthaeuser-effekt-lang}). In principle, only the step prototypes would need to be annotated with effect signatures, from which the effect type of each expression can be inferred. The knowledge that a node does not cause observable side effects during its evaluation can be leveraged for speculative execution, where a node can run even before its release time, provided it has all its required inputs. This may lead to better utilisation of the available processors while maintaining the system's deterministic functional output. The compilation scheme presented in this paper is currently being implemented as part of the Mimosa tool suite~\cite{huber-mimosa-zenodo}.

\paragraph{Acknowledgements}
 This work was partially funded by ERC through project CUSTOMER and by the Knut and Alice Wallenberg Foundation through project UPDATE.

\begingroup
\interlinepenalty=10000
\bibliographystyle{IEEEtran}
\bibliography{IEEEabrv,references}

\begin{thebibliography}{10}
\providecommand{\url}[1]{#1}
\csname url@samestyle\endcsname
\providecommand{\newblock}{\relax}
\providecommand{\bibinfo}[2]{#2}
\providecommand{\BIBentrySTDinterwordspacing}{\spaceskip=0pt\relax}
\providecommand{\BIBentryALTinterwordstretchfactor}{4}
\providecommand{\BIBentryALTinterwordspacing}{\spaceskip=\fontdimen2\font plus
\BIBentryALTinterwordstretchfactor\fontdimen3\font minus \fontdimen4\font\relax}
\providecommand{\BIBforeignlanguage}[2]{{%
\expandafter\ifx\csname l@#1\endcsname\relax
\typeout{** WARNING: IEEEtran.bst: No hyphenation pattern has been}%
\typeout{** loaded for the language `#1'. Using the pattern for}%
\typeout{** the default language instead.}%
\else
\language=\csname l@#1\endcsname
\fi
#2}}
\providecommand{\BIBdecl}{\relax}
\BIBdecl

\bibitem{wang-mimos}
W.~Yi, M.~Mohaqeqi, and S.~Graf, ``{MIMOS}: {A} {Deterministic} {Model} for the {Design} and {Update} of {Real}-{Time} {Systems},'' in \emph{Coordination Models and Languages}, M.~H. ter Beek and M.~Sirjani, Eds.\hskip 1em plus 0.5em minus 0.4em\relax Cham: Springer Nature Switzerland, 2022, pp. 17--34.

\bibitem{kahn-networks}
G.~Kahn, ``{The} {Semantics} of a {Simple} {Language} for {Parallel} {Programming},'' in \emph{Information Processing, Proceedings of the 6th {IFIP} Congress 1974, Stockholm, Sweden, August 5-10, 1974}, J.~L. Rosenfeld, Ed.\hskip 1em plus 0.5em minus 0.4em\relax North-Holland, 1974, pp. 471--475.

\bibitem{huber-mimosa}
\BIBentryALTinterwordspacing
N.~Huber, S.~Graf, P.~Rümmer, and W.~Yi, ``{Mimosa}: {A} {Language} for {Asynchronous} {Implementation} of {Embedded} {Systems} {Software},'' 2025. [Online]. Available: \url{https://arxiv.org/abs/2503.02557}
\BIBentrySTDinterwordspacing

\bibitem{biernacki-clock-directed-compilation}
D.~Biernacki, J.-L. Cola\c{c}o, G.~Hamon, and M.~Pouzet, ``Clock-directed modular code generation for synchronous data-flow languages,'' in \emph{Proceedings of the 2008 ACM SIGPLAN-SIGBED Conference on Languages, Compilers, and Tools for Embedded Systems}, ser. LCTES '08.\hskip 1em plus 0.5em minus 0.4em\relax New York, NY, USA: Association for Computing Machinery, 2008, p. 121–130.

\bibitem{caspi-lustre}
P.~Caspi, D.~Pilaud, N.~Halbwachs, and J.~A. Plaice, ``Lustre: a declarative language for real-time programming,'' in \emph{Proceedings of the 14th ACM SIGACT-SIGPLAN Symposium on Principles of Programming Languages}, ser. POPL '87.\hskip 1em plus 0.5em minus 0.4em\relax New York, NY, USA: Association for Computing Machinery, 1987, p. 178–188.

\bibitem{johnsson-lambda-lifting}
T.~Johnsson, ``{Lambda} lifting: {Transforming} programs to recursive equations,'' in \emph{Functional Programming Languages and Computer Architecture}, J.-P. Jouannaud, Ed.\hskip 1em plus 0.5em minus 0.4em\relax Berlin, Heidelberg: Springer Berlin Heidelberg, 1985, pp. 190--203.

\bibitem{damas-type-schemes}
L.~Damas and R.~Milner, ``Principal type-schemes for functional programs,'' in \emph{Proceedings of the 9th ACM SIGPLAN-SIGACT Symposium on Principles of Programming Languages}, ser. POPL '82.\hskip 1em plus 0.5em minus 0.4em\relax New York, NY, USA: Association for Computing Machinery, 1982, p. 207–212.

\bibitem{colaco-lustre-init}
J.-L. Cola{\c{c}}o and M.~Pouzet, ``Type-based initialization analysis of a synchronous dataflow language,'' \emph{International Journal on Software Tools for Technology Transfer}, vol.~6, no.~3, pp. 245--255, Aug 2004.

\bibitem{klabnik2019rust}
\BIBentryALTinterwordspacing
S.~Klabnik and C.~Nichols, \emph{The Rust Programming Language (Covers Rust 2018)}.\hskip 1em plus 0.5em minus 0.4em\relax No Starch Press, 2019. [Online]. Available: \url{https://books.google.at/books?id=0Vv6DwAAQBAJ}
\BIBentrySTDinterwordspacing

\bibitem{sml-mono}
\BIBentryALTinterwordspacing
M.~Fluet. (2025) Mlton monomorphise. [Online]. Available: \url{http://mlton.org/Monomorphise}
\BIBentrySTDinterwordspacing

\bibitem{akira-coq-mono}
A.~Tanaka, R.~Affeldt, and J.~Garrigue, ``{Safe} {Low}-level {Code} {Generation} in {Coq} {Using} {Monomorphization} and {Monadification},'' \emph{Journal of Information Processing}, vol.~26, pp. 54--72, 2018.

\bibitem{wang-mimos-tool}
W.~Yi \emph{et~al.}, ``{MIMOS} in a nutshell,'' in preparation.

\bibitem{brandes-pearl}
J.~Brandes, S.~Eichentopf, P.~Elzer, L.~Frevert, V.~Haase, H.~Mittendorf, G.~Müller, and P.~Rieder, ``{The} {Concept} of a {Process}- and {Experiment}-oriented {Programming} {Language},'' \emph{elektronische datenverarbeitung}, vol.~10, pp. 429--442, 1970.

\bibitem{burns-ravenscar}
A.~Burns, B.~Dobbing, and G.~Romanski, ``{The} {Ravenscar} tasking profile for high integrity real-time programs,'' in \emph{Reliable Software Technologies --- Ada-Europe}, L.~Asplund, Ed.\hskip 1em plus 0.5em minus 0.4em\relax Berlin, Heidelberg: Springer Berlin Heidelberg, 1998, pp. 263--275.

\bibitem{ravenscar-gnat}
J.~A. de~la Puente, J.~F. Ruiz, and J.~Zamorano, ``{An} {Open} {Ravenscar} {Real}-{Time} {Kernel} for {GNAT},'' in \emph{Reliable Software Technologies Ada-Europe 2000}, H.~B. Keller and E.~Pl{\"o}dereder, Eds.\hskip 1em plus 0.5em minus 0.4em\relax Berlin, Heidelberg: Springer Berlin Heidelberg, 2000, pp. 5--15.

\bibitem{natarajan-timed-c}
S.~Natarajan and D.~Broman, ``{Timed} {C}: {An} {Extension} to the {C} {Programming} {Language} for {Real}-{Time} {Systems},'' in \emph{2018 IEEE Real-Time and Embedded Technology and Applications Symposium (RTAS)}, 2018, pp. 227--239.

\bibitem{timed-c-compiler}
\BIBentryALTinterwordspacing
S.~Natarajan, D.~Broman, J.~Wikman, and A.~Berezovskyi, ``{KTH}'s {Timed} {C} compiler.'' [Online]. Available: \url{https://github.com/timed-c/ktc}
\BIBentrySTDinterwordspacing

\bibitem{colaco-scade}
J.-L. Cola\c{c}o, ``{An} overview of {Scade}, a synchronous language for safety-critical software (keynote),'' in \emph{Proceedings of the 7th ACM SIGPLAN International Workshop on Reactive and Event-Based Languages and Systems}, ser. REBLS 2020.\hskip 1em plus 0.5em minus 0.4em\relax New York, NY, USA: Association for Computing Machinery, 2020, p.~1.

\bibitem{berry-esterel}
G.~Berry and G.~Gonthier, ``{The} {Esterel} synchronous programming language: design, semantics, implementation,'' \emph{Science of Computer Programming}, vol.~19, no.~2, pp. 87--152, 1992.

\bibitem{benveniste-signal}
A.~Benveniste, P.~Bournai, T.~Gautier, M.~Le~Borgne, P.~Le~Guernic, and H.~Marchand, ``The {Signal} declarative synchronous language: controller synthesis and systems/architecture design,'' in \emph{Proceedings of the 40th IEEE Conference on Decision and Control (Cat. No.01CH37228)}, vol.~4, 2001, pp. 3284--3289 vol.4.

\bibitem{henzinger-giotto}
T.~A. Henzinger, B.~Horowitz, and C.~M. Kirsch, ``{Giotto}: {A} {Time}-{Triggered} {Language} for {Embedded} {Programming},'' in \emph{Embedded Software}, T.~A. Henzinger and C.~M. Kirsch, Eds.\hskip 1em plus 0.5em minus 0.4em\relax Berlin, Heidelberg: Springer Berlin Heidelberg, 2001, pp. 166--184.

\bibitem{kirsch-giotto-microkernel}
C.~M. Kirsch, M.~A.~A. Sanvido, and T.~A. Henzinger, ``A programmable microkernel for real-time systems,'' in \emph{Proceedings of the 1st ACM/USENIX International Conference on Virtual Execution Environments}, ser. VEE '05.\hskip 1em plus 0.5em minus 0.4em\relax New York, NY, USA: Association for Computing Machinery, 2005, p. 35–45.

\bibitem{henzinger-e-machine}
T.~A. Henzinger and C.~M. Kirsch, ``The embedded machine: predictable, portable real-time code,'' in \emph{Proceedings of the ACM SIGPLAN 2002 Conference on Programming Language Design and Implementation}, ser. PLDI '02.\hskip 1em plus 0.5em minus 0.4em\relax New York, NY, USA: Association for Computing Machinery, 2002, p. 315–326.

\bibitem{lohstroh-lingua-franka}
M.~Lohstroh, C.~Menard, A.~Schulz-Rosengarten, M.~Weber, J.~Castrillon, and E.~A. Lee, ``{A} {Language} for {Deterministic} {Coordination} {Across} {Multiple Timelines},'' in \emph{2020 Forum for Specification and Design Languages (FDL)}, 2020, pp. 1--8.

\bibitem{lohstroh-reactors}
M.~Lohstroh, {\'I}.~{\'I}. Romeo, A.~Goens, P.~Derler, J.~Castrillon, E.~A. Lee, and A.~Sangiovanni-Vincentelli, ``{Reactors}: {A} {Deterministic} {Model} for {Composable} {Reactive} {Systems},'' in \emph{Cyber Physical Systems. Model-Based Design}, R.~Chamberlain, M.~Edin~Grimheden, and W.~Taha, Eds.\hskip 1em plus 0.5em minus 0.4em\relax Cham: Springer International Publishing, 2020, pp. 59--85.

\bibitem{lf-gemoc}
J.~Deantoni, J.~Cambeiro, S.~Bateni, S.~Lin, and M.~Lohstroh, ``{Debugging} and {Verification} {Tools} for {Lingua} {Franca} in {Gemoc} {Studio},'' in \emph{2021 Forum on specification \& Design Languages (FDL)}, 2021, pp. 01--08.

\bibitem{bousse-gemoc}
E.~Bousse, T.~Degueule, D.~Vojtisek, T.~Mayerhofer, J.~Deantoni, and B.~Combemale, ``{Execution} framework of the gemoc studio (tool demo),'' in \emph{Proceedings of the 2016 ACM SIGPLAN International Conference on Software Language Engineering}, ser. SLE 2016.\hskip 1em plus 0.5em minus 0.4em\relax New York, NY, USA: Association for Computing Machinery, 2016, p. 84–89.

\bibitem{morel-lustre-array-iterators}
L.~Morel, ``Efficient compilation of array iterators for {Lustre},'' \emph{Electronic Notes in Theoretical Computer Science}, vol.~65, no.~5, pp. 19--26, 2002, {SLAP'2002}, Synchronous Languages, Applications, and Programming (Satellite Event of ETAPS 2002).

\bibitem{madsen-flix-lang}
M.~Madsen, ``The principles of the {Flix} programming language,'' in \emph{Proceedings of the 2022 ACM SIGPLAN International Symposium on New Ideas, New Paradigms, and Reflections on Programming and Software}, ser. Onward! 2022.\hskip 1em plus 0.5em minus 0.4em\relax New York, NY, USA: Association for Computing Machinery, 2022, p. 112–127.

\bibitem{reinking-perceus}
A.~Reinking, N.~Xie, L.~de~Moura, and D.~Leijen, ``{Perceus}: garbage free reference counting with reuse,'' in \emph{Proceedings of the 42nd ACM SIGPLAN International Conference on Programming Language Design and Implementation}, ser. PLDI 2021.\hskip 1em plus 0.5em minus 0.4em\relax New York, NY, USA: Association for Computing Machinery, 2021, p. 96–111.

\bibitem{brachthaeuser-effekt-lang}
J.~I. Brachth\"{a}user, P.~Schuster, and K.~Ostermann, ``{Effects} as capabilities: effect handlers and lightweight effect polymorphism,'' \emph{Proc. ACM Program. Lang.}, vol.~4, no. OOPSLA, Nov. 2020.

\bibitem{huber-mimosa-zenodo}
\BIBentryALTinterwordspacing
N.~Huber, ``{The} {Mimosa} {Language} - {Software} {Artifact},'' Mar. 2025. [Online]. Available: \url{https://doi.org/10.5281/zenodo.14963240}
\BIBentrySTDinterwordspacing

\end{thebibliography}
\endgroup

\end{document}